\documentclass[twocolumn,amsmath,amssymb,nofootinbib,10pt]{revtex4-2}
\usepackage{graphicx, float,amsmath, amsmath, amssymb, hyperref, subcaption}
\usepackage[braket]{qcircuit}
\usepackage{xcolor}

\begin{document}

\title{Grover's algorithm on two-way quantum computer}

\author{Grzegorz Czelusta$^{1,2}$, Dev Rishi Verma$^{3}$$^*$, Govind Wanjalkar$^{4}$$^*$}
\affiliation{
$^{1}$Institute of Theoretical Physics, Jagiellonian University, 
{\L}ojasiewicza 11, 30-348 Cracow, Poland\\
$^{2}$Doctoral School of Exact and Natural Sciences, Jagiellonian University, 
{\L}ojasiewicza 11, 30-348 Cracow, Poland\\
$^{3}$Institue of Engineering and Technology, JK Lakshmipat University, Jaipur, 302026 Rajasthan, India\\
$^{4}$Department of Physics, Indian Institute of Technology, Kharagpur, 721302 West Bengal, India\\
$^*$: Both authors contributed equally to this work
}

\date{\today}

\begin{abstract}
    Two-way quantum computing (2WQC) represents a novel approach to quantum computing that introduces a CPT version of state preparation. This paper analyses the influence of this approach on Grover's algorithm and compares the behaviour of typical Grover and its 2WQC version in the presence of noise in the system. Our findings indicate that, in an ideal scenario without noise, the 2WQC Grover algorithm exhibits a constant complexity of $\mathcal{O}(1)$. In the presence of noise, the 2WQC Grover algorithm demonstrates greater resilience to different noise models than the standard Grover's algorithm.
\end{abstract}

\maketitle

\section{Two-way quantum computers}
Standard one-way quantum computers (1WQC) assume existence of state preparation process $\ket{0}$. However, CPT symmetry of physics says that performing such process in CPT symmetry perspective, e.g. reversing used EM impulses, we should get its symmetric analog $\bra{0}$. It works similarly as postselection, but with higher success rate. This additional operation allows the construction of two-way quantum computers \cite{duda2023two} (2WQC), which are capable of more efficiently addressing NP problems by concurrently managing the flow of information.

\section{Traditional Grover}
The Grover algorithm \cite{grover1996fast} allows the identification of states that fulfil a given condition among a set of $N$ states. In the Grover algorithm, we employ two key operations: the \textit{oracle} and the \textit{diffusion}. The oracle allows us to mark states that fulfil the given condition by changing the phase of these states from a positive value to a negative one.
$$U_{Oracle}\ket{desired\;state}=-\ket{desired\;state}$$
$$U_{Oracle}\ket{other\;state}=\ket{other\;state}$$
The diffusion operator amplifies the marked states.
$$U_{Diffusion}=2H\ket{0}\bra{0}H-\mathbb{I}$$
In order to achieve the highest probability of identifying the desired states, it is necessary to repeat the oracle and diffusion approximately $\frac{\pi}{4}\sqrt{N}$. This results in a complexity of $\mathcal{O}(\sqrt{N})$, which is significantly lower than the complexity of the classical search algorithm, which is $\mathcal{O}(N)$.

\begin{figure}[h!]
    \centering
    \leavevmode
    \Qcircuit @C=1em @R=1.7em {
    &&&\mbox{Repeat $\sim \frac{\pi}{4}\sqrt{N}$}&&&\\
    & \lstick{\ket{0}} & \gate{H}& \multigate{2}{Oracle}& \multigate{2}{Diffusion}&\qw &\meter\\
    &\vdots&&&&&\vdots\\
    & \lstick{\ket{0}} & \gate{H}& \ghost{Oracle} & \ghost{Diffusion} &\qw &\meter
    \gategroup{2}{4}{3}{5}{1em}{^\}}
    }
    \caption{Typical quantum circuit for Grover}
    \label{fig:typ_grover}
\end{figure}

\subsection{Example: Sudoku solving}
\begin{figure}[h!]
    \centering
    \begin{tabular}{|c|c|}
        \hline
         $b_0$&$b_1$  \\
        \hline
         $b_2$&$b_3$ \\
        \hline
    \end{tabular}
    \quad
    \begin{tabular}{|c|c|}
        \hline
         1&0  \\
        \hline
         0&1 \\
        \hline
    \end{tabular}
    \quad
    \begin{tabular}{|c|c|}
        \hline
         1&0  \\
        \hline
         0&1 \\
        \hline
    \end{tabular}
    \caption{Binary Sudoku and its two solutions}
    \label{fig:sudoku}
\end{figure}
\begin{figure}
    \centering
    \begin{subfigure}[b]{0.45\textwidth}
    \centering
    \includegraphics[scale=0.25]{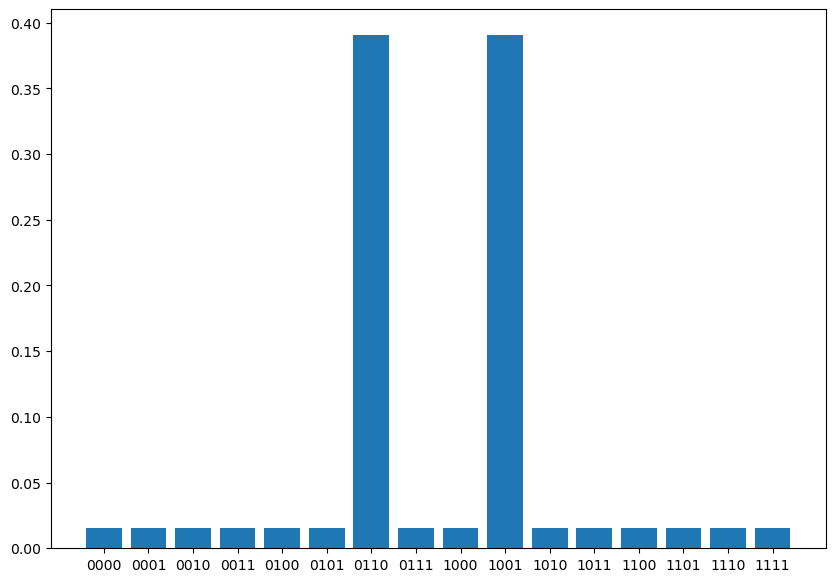}
    \caption{One repetition}
    \end{subfigure}
    \hfill
    \begin{subfigure}[b]{0.45\textwidth}
        \centering
        \includegraphics[scale=0.25]{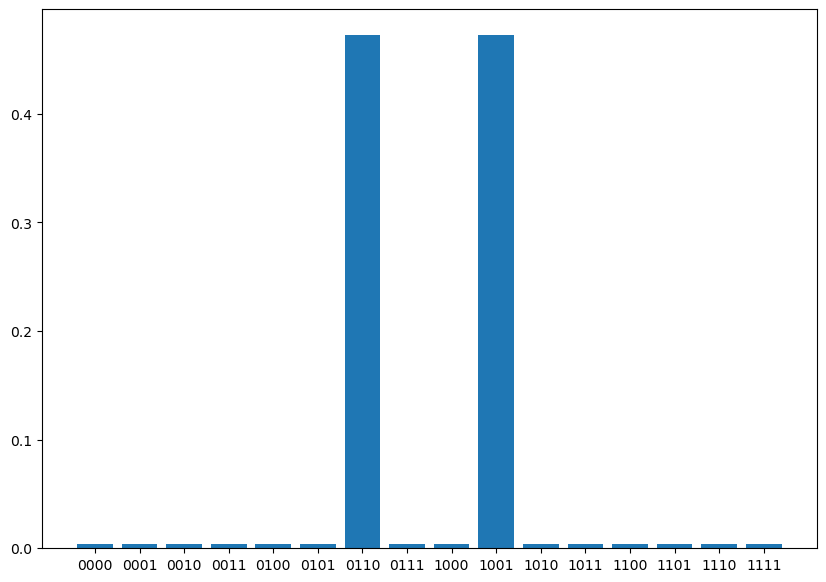}
        \caption{Two repetition}
    \end{subfigure}
    \caption{Histogram of probabilities for traditional Grover}
    \label{fig:typ_grover_hist}
\end{figure}

The Grover algorithm can be employed to identify all potential solutions to a 2x2 Sudoku puzzle Fig.\ref{fig:sudoku}. In other words, the objective is to identify strings of bits $b_0b_1b_2b_3$ that satisfy the following conditions:
$$b_0\neq b_1,\;b_0\neq b_2,\;b_1\neq b_3,\;b_2\neq b_3$$
This is equivalent to
$$b_0\oplus b_1=1,\;b_0\oplus b_2=1,\;b_1\oplus b_3=1,\;b_2\oplus b_3=1$$
As illustrated in Fig.\ref{fig:typ_grover_sudoku}, the \textit{Oracle} implementation marks the correct solutions and the diffusion process is carried out in accordance with the standard procedure.
\begin{figure}[h!]
    \centering
    \leavevmode
\Qcircuit @C=0.7em @R=1em {
     &&&&&\mbox{Oracle}&&&&&&&\mbox{Diffusion}&&&\\
     &&&&&&&&&&&&&&&\\
     & \lstick{\ket{0}} &  \gate{H} & \qw  & \ctrl{1} &  \ctrl{2} & \qw & \ctrl{2} & \ctrl{1} &\qw & \gate{H} & \gate{X}& \ctrl{1} & \gate{X}& \gate{H} & \meter\\
     & \lstick{\ket{0}} & \gate{H} & \ctrl{2}  & \targ & \qw & \ctrl{1} & \qw &  \targ & \ctrl{2}  & \gate{H} & \gate{X}& \ctrl{1} & \gate{X}& \gate{H} & \meter\\
     & \lstick{\ket{0}}& \gate{H} & \qw  & \qw &  \targ& \ctrl{1} &  \targ & \qw & \qw &  \gate{H} & \gate{X}& \ctrl{1}& \gate{X} & \gate{H} &\meter\\
     & \lstick{\ket{0}} & \gate{H} & \targ &  \qw & \qw & \ctrl{-1} & \qw & \qw & \targ & \gate{H} & \gate{X}& \ctrl{-1}& \gate{X} & \gate{H} &\meter \gategroup{3}{4}{3}{9}{3em}{^\}} \gategroup{3}{11}{3}{15}{2em}{^\}} \\
    }
    \caption{Quantum circuit for traditional Grover solving sudoku}
    \label{fig:typ_grover_sudoku}
\end{figure}
The probabilities of obtaining each state after the Grover algorithm with one or two repetitions can be observed in Fig.\ref{fig:typ_grover_hist}.

\section{2WQC Grover}
\begin{figure}[h!]
    \centering
    \includegraphics[scale=0.25]{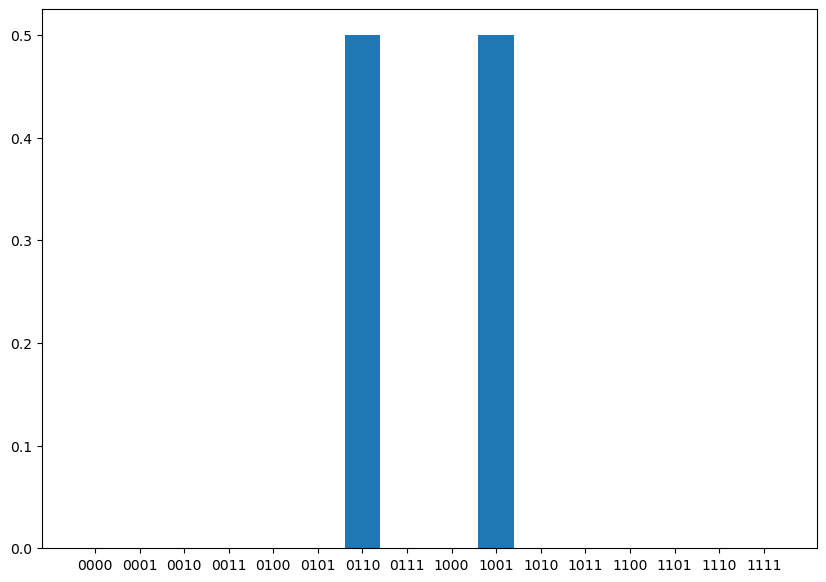}
    \caption{Histogram of probabilities for 2WQC Grover}
    \label{fig:2wqc_grover_hist}
\end{figure}

The operation $\bra{1}$ can be used in conjunction with the ancilla qubit and a modified oracle to obtain the desired states with probability equal to one, without any repetitions.
\begin{figure}[h!]
    \centering
    \leavevmode
    \Qcircuit @C=1em @R=1.7em {
    & \lstick{\ket{0}} & \gate{H}& \multigate{2}{Oracle} &\meter\\
    &\vdots&&&\vdots\\
    & \lstick{\ket{0}} & \gate{H}& \ghost{Oracle} &\meter\\
    & \lstick{\ket{0}} & \qw & \targ \qwx[-1]&\qw& \rstick{\bra{1}}
    }
    \caption{Quantum circuit for 2WQC Grover}
    \label{fig:2wqc_grover}
\end{figure}
Our new Oracle doesn't change phase of states but rather applies a $X$ gate to the ancilla qubit, which is controlled by the desired states.
$$U_{Oracle}\ket{desired\;state}\ket{0}=\ket{desired\;state}\ket{1}$$
$$U_{Oracle}\ket{other\;state}\ket{0}=\ket{other\;state}\ket{0}$$
In the context of solving the Sudoku circuit for 2WQC, Grover's method is illustrated in Fig.\ref{fig:2wqc_grover_sudoku}.
\begin{figure}[h!]
    \centering
    \leavevmode
\Qcircuit @C=1em @R=1em {
 & \lstick{\ket{0}} &  \gate{H} & \qw  & \ctrl{1} &  \ctrl{2} & \qw & \ctrl{2} & \ctrl{1} &\qw & \meter\\
 & \lstick{\ket{0}} & \gate{H} & \ctrl{2}  & \targ & \qw & \ctrl{1} & \qw &  \targ & \ctrl{2} & \meter\\
 & \lstick{\ket{0}}& \gate{H} & \qw  & \qw &  \targ& \ctrl{1} &  \targ & \qw & \qw &\meter\\
 & \lstick{\ket{0}} & \gate{H} & \targ &  \qw & \qw & \ctrl{1} & \qw & \qw & \targ &\meter\\
 & \lstick{\ket{0}} & \qw & \qw & \qw & \qw & \targ & \qw & \qw & \qw & \rstick{\bra{1}}
}
    \caption{Quantum circuit for 2WQC Grover solving Sudoku}
    \label{fig:2wqc_grover_sudoku}
\end{figure}
and the probabilities of the obtained states are shown in Fig.\ref{fig:2wqc_grover_hist}.
It is crucial to highlight that in the 2WQC Grover algorithm, the diffusion operator is not required since the correct states are automatically selected by the operation $\bra{1}$. Consequently, there is no requirement to repeat the oracle as in the case of the standard Grover's algorithm.

\section{Noise}
This study presents an analysis of the influence of different quantum noise models on both the traditional Grover algorithm and the 2WQC version of the Grover algorithm. A single-qubit error channel is applied after each gate in the circuit. The types of error channels that will be discussed are as follows.
\subsection{Bit flip}
This channel is modelled by the following Kraus matrices:
$$K_0 = \sqrt{1-p} \begin{pmatrix}
1 & 0 \\
0 & 1
\end{pmatrix}$$
$$
K_1 = \sqrt{p} \begin{pmatrix}
0 & 1 \\
1 & 0
\end{pmatrix}
$$
where $ p \in [0, 1]$  is the probability of a bit flip (Pauli $X$ error).
\begin{figure}[h!]
    \centering
    \begin{subfigure}[b]{0.45\textwidth}
    \centering
    \includegraphics[scale=0.45]{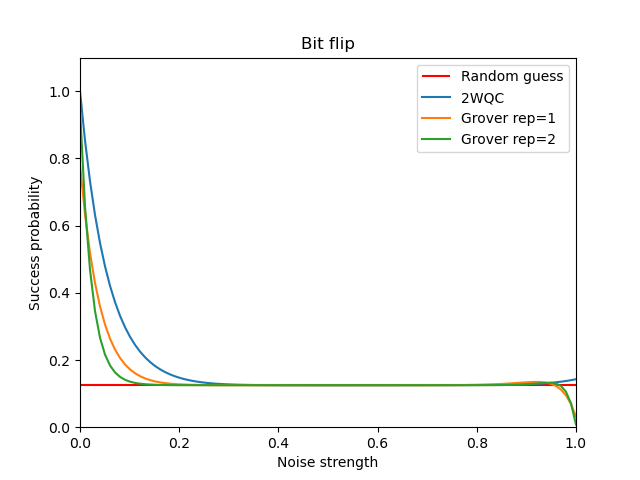}
    \end{subfigure}
    \vfill
    \begin{subfigure}[b]{0.45\textwidth}
    \centering
    \includegraphics[scale=0.45]{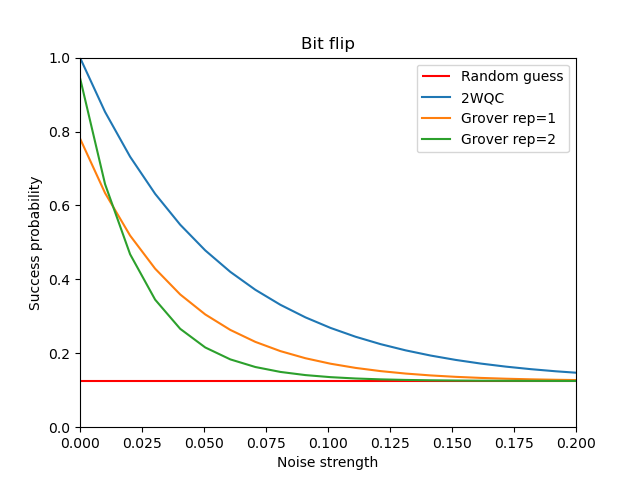}
    \end{subfigure}
    \caption{The probability of successful identification of states as a function of noise strength (in the bottom zoomed version)}
    \label{fig:BitFlip}
\end{figure}
The results, Fig.\ref{fig:BitFlip}, indicate that 2WQC Grover is more resilient to this type of noise.
\subsection{Phase flip}
This channel is modelled by the following Kraus matrices:
$$K_0 = \sqrt{1-p} \begin{pmatrix}
1 & 0 \\
0 & 1
\end{pmatrix}$$
$$K_1 = \sqrt{p} \begin{pmatrix}
1 & 0 \\
0 & -1
\end{pmatrix}$$
where $ p \in [0, 1] $ is the probability of a phase flip (Pauli $ Z $ error).

\begin{figure}[h!]
    \centering
   \includegraphics[scale=0.5]{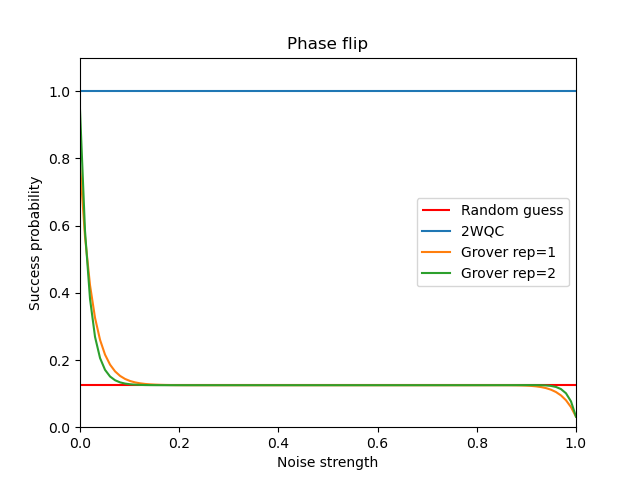}
    \caption{The probability of successful identification of states as a function of noise strength}
    \label{fig:PhaseFlip}
\end{figure}
It is somewhat surprising that the 2WQC Grover is completely resilient to this type of noise, Fig.\ref{fig:PhaseFlip}.
\subsection{Phase damping}
The interaction between the quantum system and its environment can result in the loss of quantum information without any changes in the qubit excitations. This phenomenon can be modelled by the phase damping channel, with the following Kraus matrices:
$$K_0 = \begin{pmatrix}
1 & 0 \\
0 & \sqrt{1 - \gamma}
\end{pmatrix}$$
$$K_1 = \begin{pmatrix}
0 & 0 \\
0 & \sqrt{\gamma}
\end{pmatrix}$$
where $ \gamma \in [0, 1] $ is the phase damping probability.
\begin{figure}[h!]
    \centering
   \includegraphics[scale=0.5]{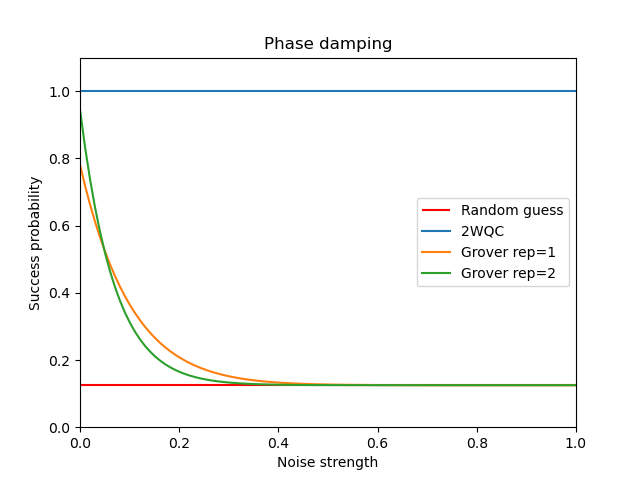}
    \caption{The probability of successful identification of states as a function of noise strength}
    \label{fig:PhaseDamping}
\end{figure}
Once more, the 2WQC Grover is entirely resilient to this type of noise, Fig.\ref{fig:PhaseDamping}.

\subsection{Depolarizing channel}
This channel is modelled by the following Kraus matrices:
$$K_0 = \sqrt{1-p} \begin{pmatrix}
1 & 0 \\
0 & 1
\end{pmatrix}$$
$$K_1 = \sqrt{\frac{p}{3}} \begin{pmatrix}
0 & 1 \\
1 & 0
\end{pmatrix}$$
$$K_2 = \sqrt{\frac{p}{3}} \begin{pmatrix}
0 & -i \\
i & 0
\end{pmatrix}$$
$$K_3 = \sqrt{\frac{p}{3}} \begin{pmatrix}
1 & 0 \\
0 & -1
\end{pmatrix}$$
where $ p \in [0, 1] $ is the depolarization probability and is equally divided in the application of all Pauli operations.

\begin{figure}[h!]
    \centering
    \begin{subfigure}[b]{0.45\textwidth}
    \centering
    \includegraphics[scale=0.45]{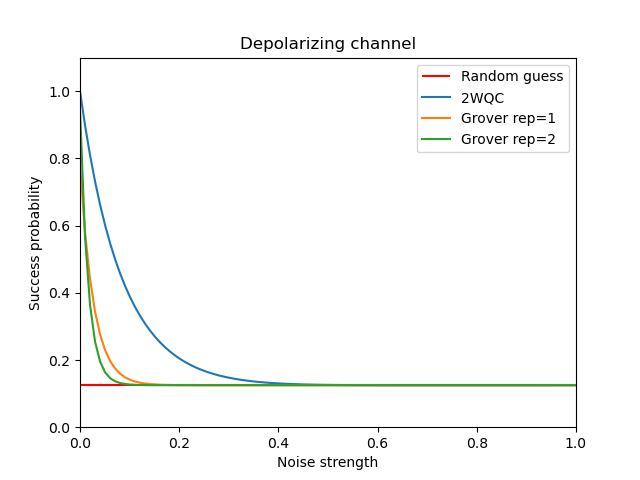}
    \end{subfigure}
    \vfill
    \begin{subfigure}[b]{0.45\textwidth}
    \centering
    \includegraphics[scale=0.45]{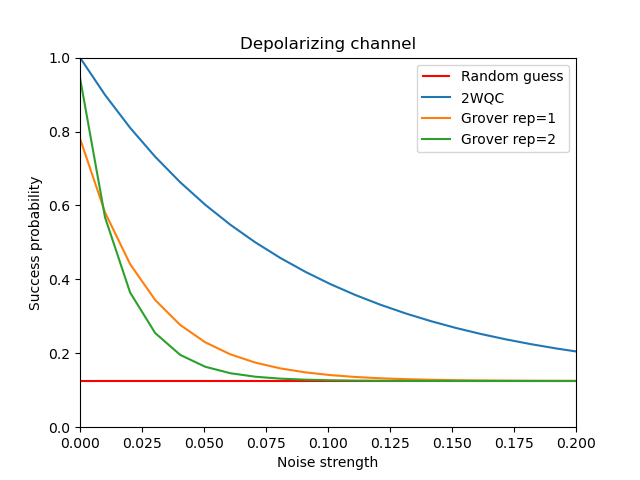}
    \end{subfigure}
    \caption{The probability of successful identification of states as a function of noise strength (in the bottom zoomed version)}
    \label{fig:DepolarizingChannel}
\end{figure}

As in the case of bit flip noise, the 2WQC Grover is more resilient to this type of noise, Fig.\ref{fig:DepolarizingChannel}.

\section{Higher numbers of qubits and noise in $\bra{1}$}
In following section we analise 2WQC Grover for higher number of qubits, in this case we don't focus on Sudoku example but our test problem is to find basis state of the form $\ket{1}^{\otimes n}$ among $n$-qubits states. Furthemore, in theoretical ideal scenario, 2WQC operation $\bra{1}$ is a perfect projection on state $\ket{1}$. Therefore, in more realistic scenario we should also consider noise in this operation.
We define imperfect 2WQC operation, with error rate $\epsilon$ as
$$\epsilon\bra{0}+(1-\epsilon)\bra{1}$$

\subsection{Circuits for 5 qubits 2WQC and 1WQC Grover Algorithm }

The 2WQC approach has been tested out for circuits upto 6 qubits Grover. Here, circuits for 5 qubits are discussed. Although, other circuits can be referred to from the GitHub repository mentioned in the references section.\cite{2WQC_Grover} 

\begin{figure}[h!]
    \centering
    \leavevmode
    \Qcircuit @C=0.8em @R=1em {
    & \lstick{\ket{0}} & \gate{H} & \ctrl{1} & \qw      & \qw      & \qw      & \qw      & \qw      & \ctrl{1} & \qw & \multigate{7}{Diffusion} & \qw & \meter \\
    & \lstick{\ket{0}} & \gate{H} & \ctrl{4} & \qw      & \qw      & \qw      & \qw      & \qw      & \ctrl{4} & \qw & \ghost{Diffusion}        & \qw & \meter \\
    & \lstick{\ket{0}} & \gate{H} & \qw      & \ctrl{1} & \qw      & \qw      & \qw      & \ctrl{1} & \qw      & \qw & \ghost{Diffusion}        & \qw & \meter\\
    & \lstick{\ket{0}} & \gate{H} & \qw      & \ctrl{3} & \qw      & \qw      & \qw      & \ctrl{3} & \qw      & \qw & \ghost{Diffusion}        & \qw & \meter\\
    & \lstick{\ket{0}} & \gate{H} & \qw      & \qw      & \qw      & \ctrl{3} & \qw      & \qw      & \qw      & \qw & \ghost{Diffusion}        & \qw & \meter\\
    & \lstick{\ket{0}} & \qw      & \targ    & \qw      & \ctrl{1} & \qw      & \ctrl{1} & \qw      & \targ    & \qw & \ghost{Diffusion}        & \qw &  \qw \\
    & \lstick{\ket{0}} & \qw      & \qw      & \targ    & \ctrl{1} & \qw      & \ctrl{1} & \targ    & \qw      & \qw & \ghost{Diffusion}        & \qw &  \qw \\
    & \lstick{\ket{0}} & \qw      & \qw      & \qw      & \targ    & \ctrl{-3} & \targ    & \qw      & \qw      & \qw & \ghost{Diffusion}        & \qw &  \qw \\
    }
    \caption{5-qubit Grover. Oracle with ancilla qubits using CZ gate to mark the state \ket{11111} \cite{vemula2022scalable}.
 }
    \label{fig:enter-label}
\end{figure}
\begin{figure}[H]
    \centering
    \leavevmode
    \Qcircuit @C=1em @R=1em {
     & \lstick{\ket{0}} & \gate{H} & \qw  & \ctrl{1} &  \qw & \multigate{4}{Diffusion} & \qw & \meter \\
     & \lstick{\ket{0}} & \gate{H} & \qw  & \ctrl{1} & \qw & \ghost{Diffusion} & \qw & \meter \\
     & \lstick{\ket{0}} & \gate{H} & \qw  & \ctrl{1} &  \qw & \ghost{Diffusion} & \qw & \meter \\
     & \lstick{\ket{0}} & \gate{H} & \qw &  \ctrl{1} & \qw & \ghost{Diffusion} & \qw & \meter \\
     & \lstick{\ket{0}} &\gate{H} & \gate{H} & \targ & \gate{H} & \ghost{Diffusion} & \qw & \meter
    }
    \caption{5-qubit Grover. Oracle without ancilla qubits using MCZ gate to mark the state \ket{11111}\cite{kumar2023noise}.}
    \label{fig:enter-label}
    
    \label{fig: 5 2wqc }
\end{figure}
\begin{figure}[H]
    \centering
    \leavevmode
    \Qcircuit @C=1em @R=1em {
    & \lstick{\ket{0}} & \gate{H} & \ctrl{5} & \qw      & \qw      & \qw      & \qw 
     & \qw      & \ctrl{5}      & \qw &  \meter \\
    & \lstick{\ket{0}} & \gate{H} & \ctrl{4} & \qw      & \qw      & \qw      & \qw 
     & \qw      & \ctrl{4}      & \qw &  \meter \\
    & \lstick{\ket{0}} & \gate{H} & \qw      & \ctrl{4} & \qw      & \qw      & \qw 
     & \ctrl{4}      & \qw      & \qw &  \meter\\
    & \lstick{\ket{0}} & \gate{H} & \qw      & \ctrl{3} & \qw      & \qw      & \qw  
     & \ctrl{3}      & \qw      & \qw & \meter\\
    & \lstick{\ket{0}} & \gate{H} & \qw      & \qw      & \qw      & \ctrl{3} & \qw  
     & \qw      & \qw      & \qw & \meter\\
    & \lstick{\ket{0}} & \qw      & \targ    & \qw      & \ctrl{1} & \qw      & \ctrl{1}   & \qw      & \targ & \qw      \\
    & \lstick{\ket{0}} & \qw      & \qw      & \targ    & \ctrl{1} & \qw      & \ctrl{1}    & \targ     & \qw & \qw      \\
    & \lstick{\ket{0}} & \qw      & \qw      & \qw      & \targ    & \ctrl{1} & \targ    & \qw          & \qw & \qw      \\
    & \lstick{\ket{0}} & \qw      & \qw    & \qw      & \qw     & \targ     & \qw   & \qw      & \qw   & \qw      & \rstick{\bra{1}}
    }
    \caption{5-qubit 2WQC Grover. Oracle with ancilla qubits to mark the state \ket{11111}.}
    \label{fig:enter-label}
\end{figure}
\begin{figure}[H]
   
    \centering
    \leavevmode
    \Qcircuit @C=1em @R=1em {
     & \lstick{\ket{0}} & \gate{H} & \qw  & \ctrl{1} &  \qw & \qw & \meter \\
     & \lstick{\ket{0}} & \gate{H} & \qw  & \ctrl{1} & \qw & \qw & \meter \\
     & \lstick{\ket{0}} & \gate{H} & \qw  & \ctrl{1} &  \qw & \qw & \meter \\
     & \lstick{\ket{0}} & \gate{H} & \qw &  \ctrl{1} & \qw & \qw & \meter \\
     & \lstick{\ket{0}} & \gate{H} & \qw &  \ctrl{1} & \qw & \qw & \meter \\
     & \lstick{\ket{0}} &\qw & \qw & \targ & \qw & \qw & \qw & \rstick{\bra{1}}
    }
    \caption{5-qubit 2WQC Grover. Oracle without ancilla qubits to mark the state \ket{11111}.}
    \label{fig:enter-label}
    \label{fig: 5 2wqc }
\end{figure}

\subsection{Results}

The results below are obtained by simulating the circuits under the condition of depolarizing channel using Pennylane\cite{bergholm2018pennylane}. However, the code for other noise models can be referred for analysis under other noise models.
The effect of noise on 2WQC and 1WQC circuits demonstrate better noise resistance in 2WQC circuits. The 2WQC without ancilla qubits shows the best results due to reduced gate-induced noise.

\begin{figure}[H]
    \centering
    \includegraphics[width=0.8\linewidth]{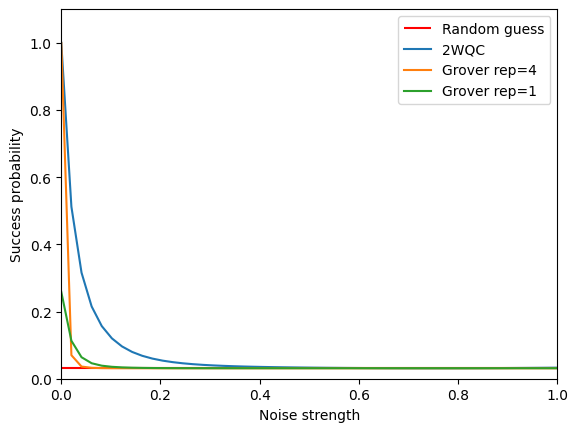}
    \caption{Noise Comparison of 2WQC Grover and 1WQC Grover with ancillary qubits.}
    \label{fig:enter-label}
\end{figure}

\begin{figure}[H]
    \centering
    \includegraphics[width=0.8\linewidth]{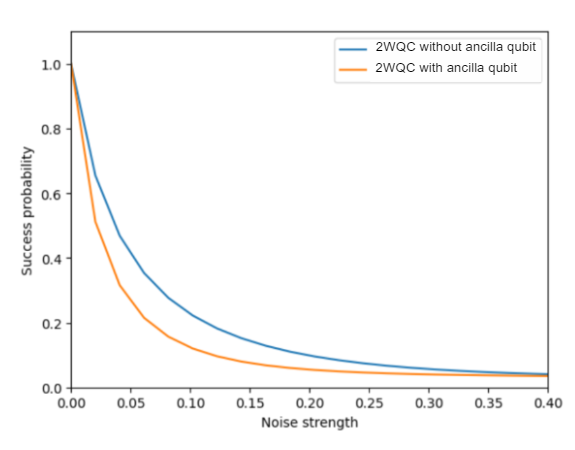}
    \caption{Noise Comparison of 2WQC Grover circuits with and without ancilla qubits.}
    \label{fig:enter-label}
\end{figure} 
\begin{figure}[H]
    \centering
    \includegraphics[width=0.8\linewidth]{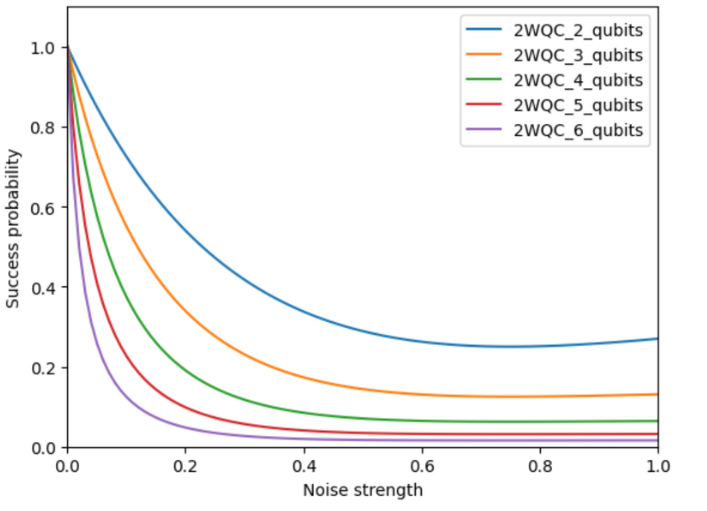}
    \caption{2WQC behaviour under noise.}
    \label{fig:enter-label}
\end{figure}
When noise is induced in \bra{1} operation, change in success probability of Sudoku is 
observed. The difference between the 2WQC Sudoku with error in \bra{1} operation and ideal 2WQC Sudoku is not significant if the error $\varepsilon$ = 0.1. Given below, the noise analysis plots for the error $\varepsilon$ = 0.2 are plotted along with ideal condition.
\begin{figure}[H]
    \centering
    \includegraphics[width=0.77\linewidth]{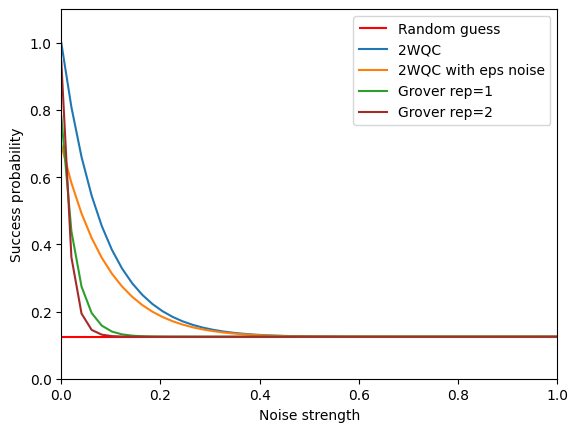}
    \caption{Sudoku circuit comparison for the depolarizing noise model with $\varepsilon$ noise in \bra{1} operation.}
    \label{fig:enter-label}
\end{figure}
\section{2WQC and postselected 1WQC}
For an $n+m$ qubit circuit, where $m$ is the number of ancilla qubits we want to postselect, the success rate of postselection is $P(c)\approx$ 2$^{-m}$. Therefore, success rate of post-selected 1WQC ($S_{1WQC}$) algorithm and success rate of 2WQC ($S_{2WQC}$) algorithm are related by \cite{duda2024three}:
\begin{equation}
    S_{1WQC} = S_{2WQC} \cdot S_{PS}
\end{equation}
where $S_{PS}$ is the success rate of post-selection.

Now, since the probability for 2WQC success rate is nearly 1 (in the almost ideal case) then the above formula reduced to:
\begin{equation}
    S_{1WQC} \approx S_{PS}
    \nonumber
\end{equation}
We see that succes rate of postselection is a bottle-neck for success rate of 1WQC. Example of this behaviour can be found in GitHub repository \cite{2WQC_Grover}.


\section{Methods}
All presented results were obtained in simulations using the PennyLane \cite{bergholm2018pennylane} and NumPy \cite{harris2020array} Python libraries. A code is provided in the GitHub repositories \cite{repo, 2WQC_Grover} to recreate these results and to allow users to perform their own simulations.

\section{Summary}
The implementation of the 2WQC operation allows for the modification of the Grover algorithm, which now runs in a constant time complexity of $\mathcal{O}(1)$, in contrast to the standard algorithm, which runs in a time complexity of $\mathcal{O}(\sqrt{N})$. Moreover, it is more resilient to the various types of noise. In the case of phase flip and phase damping channels, the system is completely resilient. In the absence of noise, the 2WQC Grover algorithm is capable of identifying all states with absolute certainty, in contrast to the traditional Grover algorithm, which is only capable of identifying states with some (high) probability.

\bibliography{bibliography.bib}

\end{document}